\documentclass[aps,prl,showpacs,twocolumn,superscriptaddress]{revtex4-1}
\usepackage{bm,color}
\usepackage{graphicx}
\usepackage{amsmath,amssymb}
\usepackage{braket}
\begin{document}
\title{Thermoelectric Effect of Skyrmion Crystal Confined in a Magnetic Disk}
\author{Junnosuke Matsuki}
\affiliation
{Department of Applied Physics, Waseda University, Okubo, Shinjuku-ku, Tokyo 169-8555, Japan}
\affiliation
{Department of Basic Science, The University of Tokyo, Komaba, Meguro, Tokyo 153-8902, Japan}
\author{Masahito Mochizuki}
\affiliation
{Department of Applied Physics, Waseda University, Okubo, Shinjuku-ku, Tokyo 169-8555, Japan}
\begin{abstract}
We theoretically propose that an electric voltage can be generated by thermal gradient with a rotating skyrmion crystal confined in a magnetic disk. We find that the rotation of skyrmion crystal induced by diffusive thermal magnon currents in the presence of temperature gradient gives rise to spinmotive forces in the radial direction through coupling to conduction-electron spins. The amplitude of generated spinmotive force is larger for a larger temperature gradient at a lower temperature. The proposed phenomenon can be exploited as spintronics-based thermoelectric devices to realize the conversion of heat to electricity.
\end{abstract}
\maketitle

Noncollinear magnetic structures such as domain walls~\cite{Hubert98}, magnetic vortices~\cite{Shinjo00,Wachowiak02,Raabe00,Yamada07}, magnetic spirals~\cite{Nagaosa19,Yokouchi20,Kitaori20,Kurebayashi21}, and skyrmions~\cite{Bogdanov89,Bogdanov94,Rossler06,Nagaosa13,Mochizuki15,Seki16,Finocchio16,Everschor19,Tokura21} often give rise to intriguing transport phenomena through coupling to conduction electrons~\cite{MaekawaB06,SeidelB16,MaekawaB17}. The spinmotive force, i.e., the emergent electric field induced by magnetization dynamics, is one of the most important examples~\cite{Volovik87,Barnes07}. Noncollinear magnetizations produce spatially varying effective vector potentials acting on conduction electrons through exchange coupling. The magnetization dynamics causes their temporal variation and eventually induces effective electromotive forces that act on the conduction electrons. The spinmotive force can be interpreted as an inverse effect of the spin-transfer torque~\cite{MaekawaB17,Maekawa13}, that is, injection of spin-polarized electric currents drive noncollinear magnetic structures via angular-momentum transfer from conduction-electron spins to the noncollinear magnetizations~\cite{MaekawaB06,Slonczewski96,Yamaguchi06,Parkin08,Yamanouchi04,Tatara04,Yamanouchi06}. Several experimental observations~\cite{SAYang09,PNHai09,Yamane11,Hayashi12,Tanabe12,Zhou19,Fukuda20} and theoretical proposals~\cite{Barnes06,Ohe09a,Ohe09b,Kishine12,KWKim12,Ohe13,Shimada15,Koide19,Yamane19} have been reported so far. The spinmotive force is formulated as~\cite{Volovik87},
\begin{eqnarray}
E_\mu(\bm r, t)=\frac{\hbar}{2e}\bm m \cdot 
(\partial_\mu \bm m \times \partial _t \bm m)
\quad \quad (\mu=x, y).
\label{eq:smf1}
\end{eqnarray}
Here $\bm m(\bm r, t)$ is the normalized classical magnetization vector. This formula implies that magnetizations varying both spatially and temporally can induce the spinmotive force.

The magnetization dynamics required to generate the spinmotive force can be realized by several ways. Simple application of static magnetic field can drive translational motion of domain walls in ferromagnetic nanowires, which results in the generation of direct-current (DC) spinmotive forces. Application of microwave magnetic field can drive the gyrotropic motion of magnetic vortex confined in a disk-shaped magnet. It is possible to achieve enhanced dynamics of magnetic vortex through tuning the microwave frequency at a resonance frequency of the system, which generates the alternate-current (ac) spinmotive forces. Magnetic skyrmions in chiral magnets can be excited by microwave application~\cite{Mochizuki12,Petrova11,Mochizuki15,Schwarze15,Garst17,Bo22}, and the generation of spinmotive forces with the microwave-induced skyrmion dynamics was theoretically predicted recently~\cite{Ohe13,Shimada15,Koide19,Yamane19}.

It was discovered that a skyrmion crystal~\cite{Bogdanov89,Bogdanov94,Muhlbauer09,XZYu10}, i.e.,  a crystallized assembly of magnetic skyrmions confined in a thin-plate magnet exhibits rotational motion when a radial temperature gradient is present~\cite{Mochizuki14,Everschor12}. This special motion of skyrmion crystal is induced by the topological magnon Hall effect~\cite{Mochizuki14}. More specifically, thermally activated oscillations of magnetizations propagate from the center to the peripheries of the sample along the temperature gradient. These thermal diffusion currents of magnetization oscillations, i.e., thermal magnon currents, are suffered from emergent magnetic fields generated by the topological skyrmion textures~\cite{Nagaosa12,ZangJ11,Schulz12}, and eventually the propagation directions are deflected. The reaction force of this effect render the skyrmion crystal rotate within the specimen in a direction opposite to the magnon-current deflection. 

\begin{figure}[tb]
\centering
\includegraphics[scale=0.5]{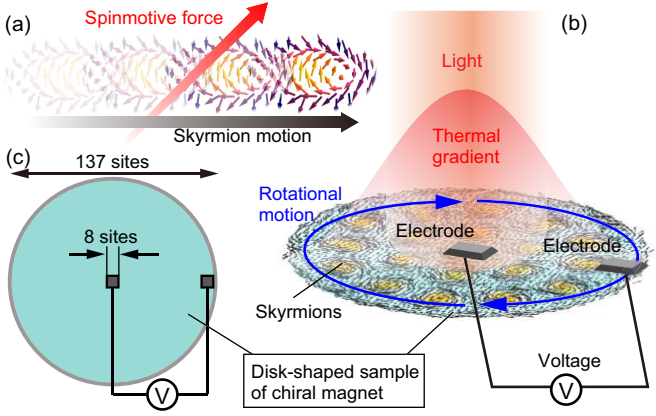}
\caption{(a) A moving skyrmion generates a spinmotive force perpendicular to its moving direction. (b) Schematics of the thermoelectric voltage generation with a skyrmion crystal confined in a nanodisk. A thermal gradient induced by irradiation gives rise to rotational motion of the skyrmion crystal, which generates spinmotive forces in radial directions. (c) Setup of the numerical simulations.}
\label{Fig01}
\end{figure}

In this Letter, we theoretically propose that this thermally induced rotational motion of skyrmions can be exploited for the  DC-voltage generation. The radial temperature gradient can be realized by irradiating the system with light or electron beam. According to the formula in Eq.~(\ref{eq:smf1}), a moving skyrmion texture can generate a spinmotive force perpendicular to its moving direction [Fig.~\ref{Fig01}(a)]. This indicates that the rotational motion of skyrmions confined in a circular disk generates an DC electric voltage in the radial direction as shown in Fig.~\ref{Fig01}(b). We show that the expected electric voltage is large enough to observe experimentally. The proposed phenomenon can be exploited as spintronics-based thermoelectric devices to convert thermal energies to electric powers. A thermoelectric effect of magnetic skyrmions was previously discussed in Ref.~\cite{Hoshi20}, and its microscopic mechanism is a magnon-drag effect of static skyrmions~\cite{Hoshi20}. Our proposal based on the spinmotive force generated by thermally driven skyrmions is distinct from this previous proposal.

We use a classical Heisenberg model on a square lattice, which contains the ferromagnetic-exchange interactions, the Dzyaloshinskii-Moriya (DM) interactions~\cite{Dzyaloshinsky58,Moriya60a,Moriya60b}, and the Zeeman coupling to an external magnetic field $\bm B=(0,0,B_z)$ normal to the plane. The Hamiltonian is given by~\cite{Bak80,YiSD09},
\begin{eqnarray}
\mathcal{H}&=&-J \sum_{<i,j>} \bm{m}_i \cdot \bm{m}_j
- \bm{B} \cdot \sum_i \bm{m}_i
\nonumber \\
& &-D \sum_i \left( \bm{m}_i \times \bm{m}_{i+ \hat{\bm{x}}} \cdot \hat{\bm{x}} + \bm{m}_i \times \bm{m}_{i + \hat{\bm{y}}} \cdot \hat{\bm{y}} \right).
\label{hamiltonian}
\end{eqnarray}
The magnetization vector $\bm m_i$ is defined as $\bm m_i=-\bm S_i/\hbar$ with $\bm S_i$ being the classical spin vector. We adopt $J=1$ as the energy units and set $D/J=0.27$. The strength of the external magnetic field is fixed at $B_z/J=-0.03$, which corresponds to $\sim$0.26 T. The unit conversions when $J=1$ meV are summarized in Table~\ref{unitconversion}.
\begin{table}[tb]
\caption{Unit conversion table for $J=1$ meV.}
\label{unitconversion}
\centering
\begin{tabular}{lcc}
\hline \hline
 &  
\begin{tabular}{c}
Dimensionless \\ quantity
\end{tabular}
 &  
\begin{tabular}{c}
Corresponding value \\ with units
\end{tabular} \\
\hline
Exchange interaction & $J=1$ & $J=1$ meV \\
Time & $\tau=1$  & $\hbar/J = 0.66$ ps \\
Temperature & $T=1$ & $J/k_{\rm B}=11.6$ K \\
\hline \hline
\end{tabular}
\end{table}

In this study, we treat a skyrmion crystal confined in a disk-shaped specimen of chiral magnet with diameter of $N_{\rm R}$=137 sites. We impose the open boundary condition and simulate thermally induced dynamics of this skyrmion microcrystal by numerically solving the stochastic Landau-Lifshitz-Gilbert (sLLG) equation using the Heun scheme. The sLLG equation is given by,
\begin{equation}
\begin{split}
\frac{d\bm m_i}{d\tau}=&
-\frac{1}{1+\alpha_{\rm G}^2}\left[
\bm m_i \times \left(\bm B^{\rm eff}_i + \bm \xi_i(\tau) \right) \right. \\
& \left. 
+\frac{\alpha_{\rm G}}{m} \bm m_i \times \left\{
\bm m_i \times \left(\bm B^{\rm eff}_i + \bm \xi_i(\tau) \right) \right\} 
\right].
\end{split}
\end{equation}
Here $\alpha_{\rm G}$ is the Gilbert-damping coefficient, which is fixed at a typical value of $\alpha_{\rm G}=0.01$ throughout the following calculations, and $\bm B^{\rm eff}_i=-\partial\mathcal{H}/\partial\bm m_i$ is the deterministic field derived from the Hamiltonian in Eq.~(\ref{hamiltonian}). The Gaussian stochastic field $\bm \xi_i(\tau)$ describes the effects of thermally fluctuating environment interacting with the magnetizations $\bm m_i$, which satisfies $\langle \xi_{i,\mu}(\tau)\rangle =0$ and $\langle \xi_{i,\mu}(\tau) \xi_{j,\nu}(\tau')\rangle =2\kappa_i \delta_{ij} \delta_{\mu \nu} \delta(\tau-\tau')$ with $\mu$ and $\nu$ are the Cartesian indices. The fluctuation-dissipation theorem gives a relation between $\kappa_i$ and temperature $T_i$ as $\kappa_i=\alpha_{\rm G}T_i/m$. Here $T_i$ is the local temperature at the $i$th site, and we assume a linear temperature gradient as,
\begin{equation}
T_i=T_0 + \frac{r_i}{R}\Delta T,
\end{equation}
where $T_0$ and $T_0+\Delta T$ are temperatures at center and periphery of the circular disk, respectively, and $r_i$ and $R \equiv a(N_{\rm R}-1)/2$ are the distance of the $i$th site from the center and the radius of the disk, respectively. We assume the lattice constant of $a$=0.5 nm, for which $\Delta T$=0.01 corresponds to the temperature gradient of $\sim$0.34 K/$\mu$m. The initial magnetization configuration of skyrmion crystal is prepared by the Monte Carlo thermalization at low temperatures and by further relaxing them by numerical simulations using the Landau-Lifshitz-Gilbert equation at $T_0=0$ and $\Delta T=0$. 

Using the simulated spatiotemporal profiles of magnetizations, we calculate the spinmotive force $\bm E_i(\tau)$ induced by the thermal skyrmion dynamics. For the numerical calculation, we use the discretized form of Eq.~(\ref{eq:smf1}),
\begin{align}
E_{\mu, i}=
&\frac{\hbar}{2e} \bm{m}_i(\tau) \cdot \left( \frac{\bm{m}_{i+ \hat{\mu}}(\tau) - \bm{m}_{i-\hat{\mu}}(\tau)}{2a} \right. \nonumber \\
& \quad \left. \times \frac{\bm{m}_i(\tau+\Delta \tau) - \bm{m}_i(\tau - \Delta \tau)}{2 \Delta \tau} \right),
\label{eq:smf2}
\end{align}
where $\mu$(=$x$, $y$) is the Cartesian indices. Using thus calculated spatiotemporal profiles of the spinmotive forces $\bm E_i(\tau)$, we calculate the spatiotemporal profiles of electric potential $\phi_i(\tau)$ by solving the Poisson equation $\nabla^2 \phi_i=-\nabla \cdot \bm E_i$. We assume two electrodes attached on the system as shown in Fig.~\ref{Fig01}(c). The electric potential $\phi_{\rm e}$ at each electrode is defined by 
\begin{equation}
\phi_{\rm e} = \frac{1}{N_{\rm e}}\sum_{i \in \text{electrode area}} \phi_i,
\end{equation}
where $N_{\rm e}$ is the number of sites within the area designated as the electrode.

First the micromagnetic simulations using the sLLG equation have confirmed that the skyrmion crystal confined in the magnetic disk start rotating in the presence of radial temperature gradient as discussed in the previous study~\cite{Mochizuki14}, whereas the skyrmions constituting the skyrmion crystal exhibit Brownian motion only when the temperature gradient is absent. We have also revealed that the skyrmion crystal rotates more quickly for larger temperature gradient $\Delta T$ and lower temperature $T_0$.

\begin{figure}[tb]
\centering
\includegraphics[scale=1.0]{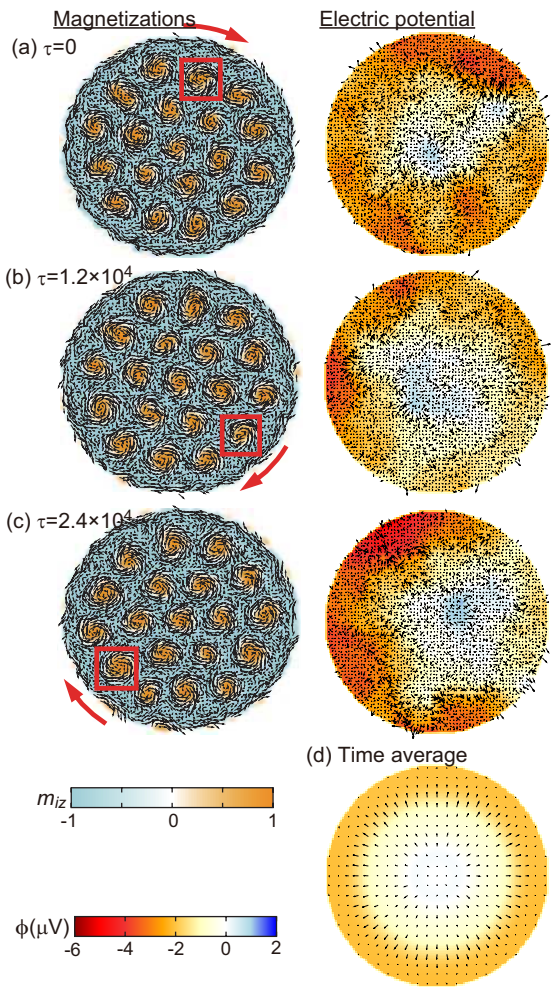}
\caption{(Left panels) Snapshots of the dynamical magnetizations $\bm m_i(\tau)$ of skyrmion crystal rotating in a circular disk in the presence of temperature gradient $\Delta T$=0.04 when $T_0=0.10$ at selected moments, i.e., (a) $\tau=0$, (b) $\tau=1.2 \times 10^4$, and (c) $\tau=2.4 \times 10^4$. In-plane components of magnetizations are represented by arrows, while out-of-plane components are represented by colors. The skyrmion crystal is rotating in a clockwise sense. One skyrmion marked by solid squares is focused on to trace a trajectory of the rotational motion. (Right panels) Spatial distributions of the electric potential $\phi_i(\tau)$ at the corresponding moments. Arrows indicate the local electric-field vectors. (d) Spatial distribution of the time-averaged electric potential $\bar{\phi_i(\tau)}$. In (d), the lengths of the arrows, which are proportional to the local electric-field strengths, are drawn on a scale six times larger than in (a)-(c).}
\label{Fig02}
\end{figure}
Left panels of Figs.~\ref{Fig02}(a)-(c) show snapshots of the simulated dynamics of magnetizations $\bm m_i(\tau)$ for rotating skyrmion crystal in the presence of temperature gradient $\Delta T$=0.04 at $T_0=0.1$ at selected moments, i.e., (a) $\tau=0$, (b) $\tau=1.2 \times 10^4$, and (c) $\tau=2.4 \times 10^4$. Here the arrows represent the in-plane components of magnetization vectors, and the colors represent their out-of-plane components. One skyrmion marked by solid squares is focused on to trace a trajectory of the rotational motion, which indicates that the skyrmion crystal exhibits unidirectional rotation in a clockwise sense. Right panels of Figs.~\ref{Fig02}(a)-(c) show snapshots of the calculated spatial distributions of electric potential $\phi_i(\tau)$ at the corresponding moments, where we set $\phi_i(\tau)=0$ at the center of the disk. The potential near the center is higher than that at the periphery. We also find that although it is slightly asymmetric because of thermal fluctuations, the spatial profile of the electric potential is nearly concentric as can be clearly seen in its time average [Fig.~\ref{Fig02}(d)].

\begin{figure}[tb]
\centering
\includegraphics[scale=1.0]{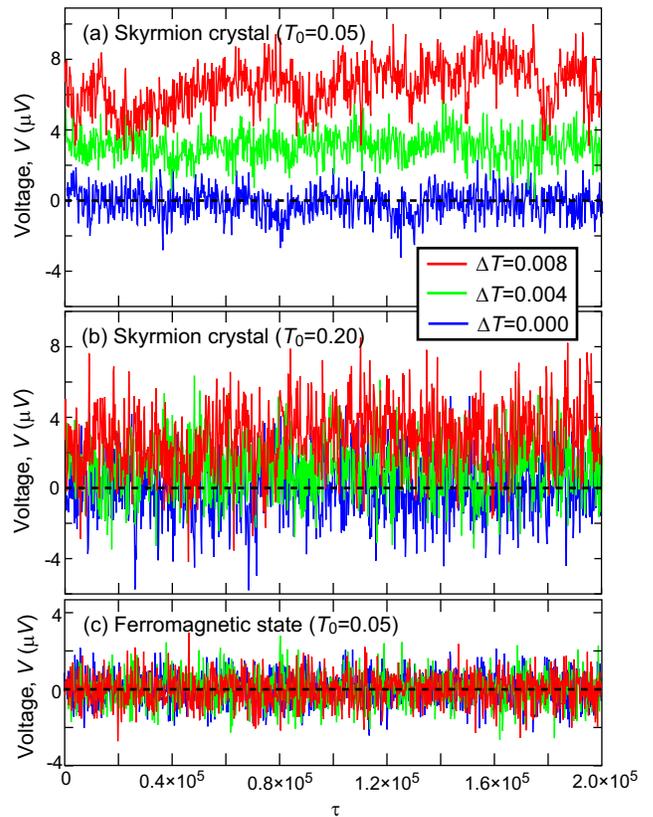}
\caption{(a), (b) Simulated time profiles of the electric voltage generated by the rotating skyrmion crystal confined in a magnetic disk for various magnitudes of temperature gradient $\Delta T$ at fixed temperatures (a) $T_0$=0.05 and (b) $T_0$=0.2. (c) Those generated in a ferromagnetic state at $T_0$=0.05.}
\label{Fig03}
\end{figure}
Next we study the time profiles of electric voltage between the two electrodes mounted on the center and edge of the magnetic disk. Figures~\ref{Fig03}(a) and (b) show the simulated time profiles of the generated electric voltage $V(\tau)$ between the two electrodes for various magnitudes of temperature gradients, i.e., $\Delta T$=0, 0.004, and 0.008 at different temperatures (a) $T_0$=0.05 and (b) $T_0$=0.2. We find that when the temperature gradient is absent (i.e., $\Delta T$=0), the electric voltage is fluctuating around zero, indicating zero electric voltage in time average. On the contrary, with the nonzero temperature gradients (i.e., $\Delta T$=0.004, 0.008), the voltages are again fluctuating, but the fluctuations occur noticeably around nonzero values, indicating the appearance of nonzero electric voltage in time average. We also find that a larger electric voltage is observed at lower temperatures in comparison between the voltage profiles for the same temperature gradient $\Delta T$ at different temperatures, i.e., (a) $T_0$=0.05 and (b) $T_0$=0.20. We have also examined the electric voltage for a pure ferromagnetic case without skyrmions for comparison as shown in Fig.~\ref{Fig03}(c). We find that the electric voltages are fluctuating around zero in the time profiles and their time-averaged value is zero even in the presence of temperature gradient. 

\begin{figure}[tb]
\centering
\includegraphics[scale=1.0]{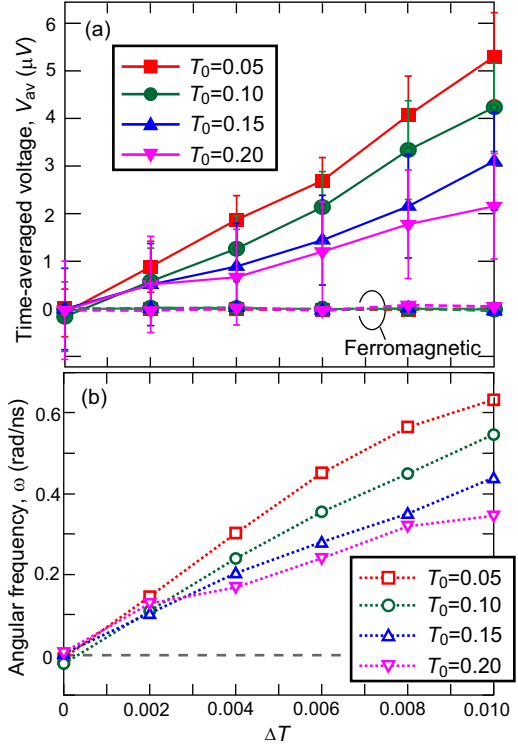}
\caption{(a) Simulated time-averaged electric voltages generated by the rotating skyrmion crystal. Those in the ferromagnetic state are also shown. (b) Simulated time-averaged angular velocities of the skyrmion-crystal rotation. They are plotted as functions of the temperature gradient $\Delta T$ at several temperatures $T_0$.}
\label{Fig04}
\end{figure}
In Figs.~\ref{Fig04}(a) and (b), we plot the simulated time-averaged electric voltages $V_{\rm av}$  and the time-averaged angular velocities $\omega$ of the rotation, respectively, as functions of the temperature gradient $\Delta T$ at various temperatures $T_0$. We first find in Fig.~\ref{Fig04}(a) that the generated electric voltage increases as the temperature gradient $\Delta T$ increases when $T_0$ is fixed. In addition, when $\Delta T$ is fixed, a larger electric voltage is observed at lower $T_0$. These tendencies coincide well with the behaviors of $\omega$ in Fig.~\ref{Fig04}(b) indicating that the dependencies of generated electric voltage on $\Delta T$ and $T_0$ are governed by the those of the rotation speed of skyrmion crystal. At higher $T_0$, the thermal fluctuations of emergent field are more significant, which results in the decrease of the rotation speed and eventually the suppression of the voltage.

The generated spinmotive force is closely related with the topological charge density of magnetic textures. Given that the magnetic texture is hardly deformed during the motion, the magnetization vectors $\bm m(\bm r, \tau)$ constituting the magnetic texture are determined by the relative coordinates $\bm \xi(\tau) \equiv \bm r - \bm R(\tau)$ only as,
\begin{equation}
\bm m(\bm r, \tau) = \bm m(\bm \xi(\tau)) = \bm m(\bm r - \bm R(\tau)),
\end{equation}
where $\bm R(\tau)$ is the center-of-mass coordinates of the magnetic texture. Then we obtain,
\begin{align}
\partial_t \bm m(\bm r - \bm R(\tau))
= \sum_\nu \frac{\partial \bm m}{\partial \xi_\mu}\partial_t \xi_\nu
=-\sum_\nu \frac{\partial \bm m}{\partial \xi_\mu}\partial_t R_\nu.
\end{align}
Using this relationship, Eq.~(\ref{eq:smf1}) can be rewritten as,
\begin{equation}
E_\mu=-\frac{\hbar}{2e} \bm m \cdot 
\left(\partial_\mu \bm m \times \partial_\nu \bm m \right)v_\nu,
\label{eq:toposmf}
\end{equation}
where $v_\nu$ $(\nu=x, y)$ is the velocity of magnetic texture. This equation contains the topological charge density and indicates that a moving topological magnetic texture generates a spinmotive force perpendicular to its motion. Accordingly, a rotating skyrmion crystal generates a spinmotive force in radial directions from the center to periphery of disk-shaped specimen. The spatial derivative $\partial_\mu \bm m$ in the equation indicates that smaller sized skyrmions with more rapidly varying magnetizations are appropriate for the larger voltage generation.

In summary, we have theoretically demonstrated that continuous unidirectional rotation of skyrmion crystal confined in a magnetic disk, which occurs in the presence of radial temperature gradient, gives rise to persistent nonzero DC spinmotive force in the radial direction. Consequently, nonzero electric voltage appears between electrodes set at the center and edge of the disk. In the present work, we employ a very small-sized system of 68-nm-diameter disk containing 19 skyrmions to reduce computational costs of the numerical simulations. 
Because each skyrmion in skyrmion crystals behaves as a battery in series connection, we expect a larger voltage for a larger skyrmion crystal if we could properly introduce the thermal gradient. An advantage of the usage of skyrmion crystal in a circular disk is that we can realize the rotational motion with thermal gradient, which enables the persistent generation of DC voltage, whereas in the stripline system with a linear temperature gradient, only the translational motion is possible, which inevitably stops when they reach the system edge, and thus only the generation of pulse or AC voltage is possible.  We also note that a similar DC-voltage generation by application of thermal gradient to a confined skyrmion crystal was proposed previously~\cite{Hoshi20}. Its physical mechanism is a phenomenon called magnon drag, which is different from our proposal. Importantly, their effect can be expected for pinned skyrmion crystals without motion, whereas our effect takes place when the skyrmion crystal rotates. This difference might give us an opportunity to compare the contributions from different origins experimentally.

We have also revealed that the induced electric voltage depends on the temperature gradient as well as the temperature of the disk, that is, a larger voltage is obtained at a lower temperature with a larger temperature gradient. This indicates that the induced electric voltage can be controlled by tuning the intensity of light irradiating the disk to adjust the temperature and the temperature gradient. Through these demonstrations, we have proposed an efficient method to convert heat to electricity using skyrmion-hosting magnets, which might be exploited for future spintronics-based thermoelectric devices. It is also worth mentioning that this phenomenon can be expected for other topological magnetic textures such as merons and hedgehogs, since the magnitude of induced electromotive force is governed by the topological charge density in the disk. Further studies on the proper device form are required to achieve the high energy efficiency, the advanced heat control, and the sophisticated architectures towards practical applications of our proposal.

\section{Acknowledgement}
This work was supported by 
JSPS KAKENHI (Grant No.~20H00337), 
JST CREST (Project No.~JPMJCR20T1),
and Waseda University Grant for Special Research Projects (No.~2022C-139).


\end{document}